\documentclass[11pt]{article}
\pdfoutput=1

\usepackage{blois}

\def\be{\begin{equation}}
\def\ee{\end{equation}}
\def\bea{\begin{eqnarray}}
\def\eea{\end{eqnarray}}

\newcommand{\Photo}{}

\begin{document}

\rightline{KCL-PH-TH/2013-31}

\vspace*{3cm}
\title{Constraining Higgs Properties in Associated Production\footnote{Talk given on May 2013 at the $25^{th}$ \emph{Rencontres de Blois}.} }

\author{ Tevong~You }

\address{Theoretical Particle Physics and Cosmology Group, Physics Department, \\
King's College London, London WC2R 2LS, UK}

\maketitle\abstracts{
The Higgs boson $H$ produced in association with a vector boson $V=W^\pm, Z$ is the main production mechanism for searches at the Tevatron and forms an important part of Higgs analyses at the LHC. We show here that the $V+H$ invariant mass distribution and the energy dependence of associated production provide powerful ways of constraining Higgs properties, such as its spin-parity and dimension-6 operator coefficients. 
}

\section{Introduction}

Following the historic discovery of a Higgs boson at the LHC~\cite{discovery}, accompanied by strong evidence from the Tevatron~\cite{evidence}, attention now turns towards the question whether it is indeed \emph{the} Higgs particle of the Standard Model (SM). The question is:~to what extent does the current data allow us to infer that this is a spin-zero elementary scalar with even parity responsible for breaking the electroweak symmetry? 

It was first noted by Miller et al.~\cite{MCEMZ} that the $e^+ e^- \to V + X$ reaction would be sensitive to the spin-parity of the $X$ due to the different threshold behaviour of the $V+X$ invariant mass distribution. In Section \ref{sec:invmass} we consider this process at hadron colliders and find that the discriminating power of associated production remains after simulating typical cuts at the LHC and Tevatron. In particular the D0 experiment at the latter is already able to exclude a spin two hypothesis at $99.9\%$ CL using this method in the $H\to b\bar{b}$ channel alone~\cite{D0}. In Section \ref{sec:energy} the point is made that the observation of a signal at the Tevatron and LHC already strongly exclude spin two, as well as placing competitive limits on dimension-6 operator coefficients. This is due to the different energy dependence of associated production for the various hypotheses. We conclude in Section \ref{sec:conclusion}. 

The work on which this contribution to the conference proceedings was based on can be found in Ellis et al.~\cite{fasttrack}~\cite{secundafacie}.

\section{$V+X$ Invariant Mass Distribution}
\label{sec:invmass}

\begin{figure}[h!]
\centering
\includegraphics[scale=0.27]{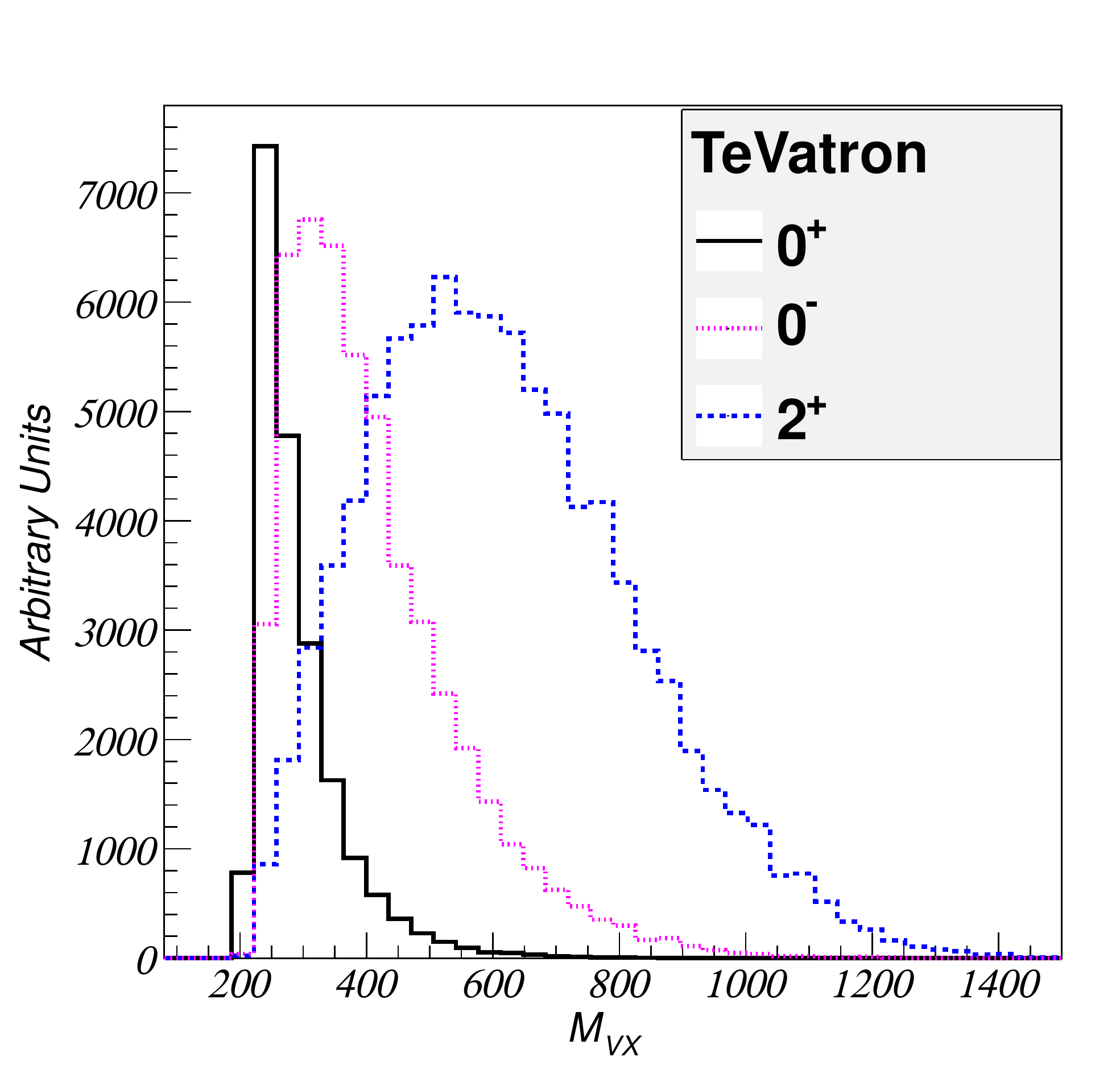}
\includegraphics[scale=0.27]{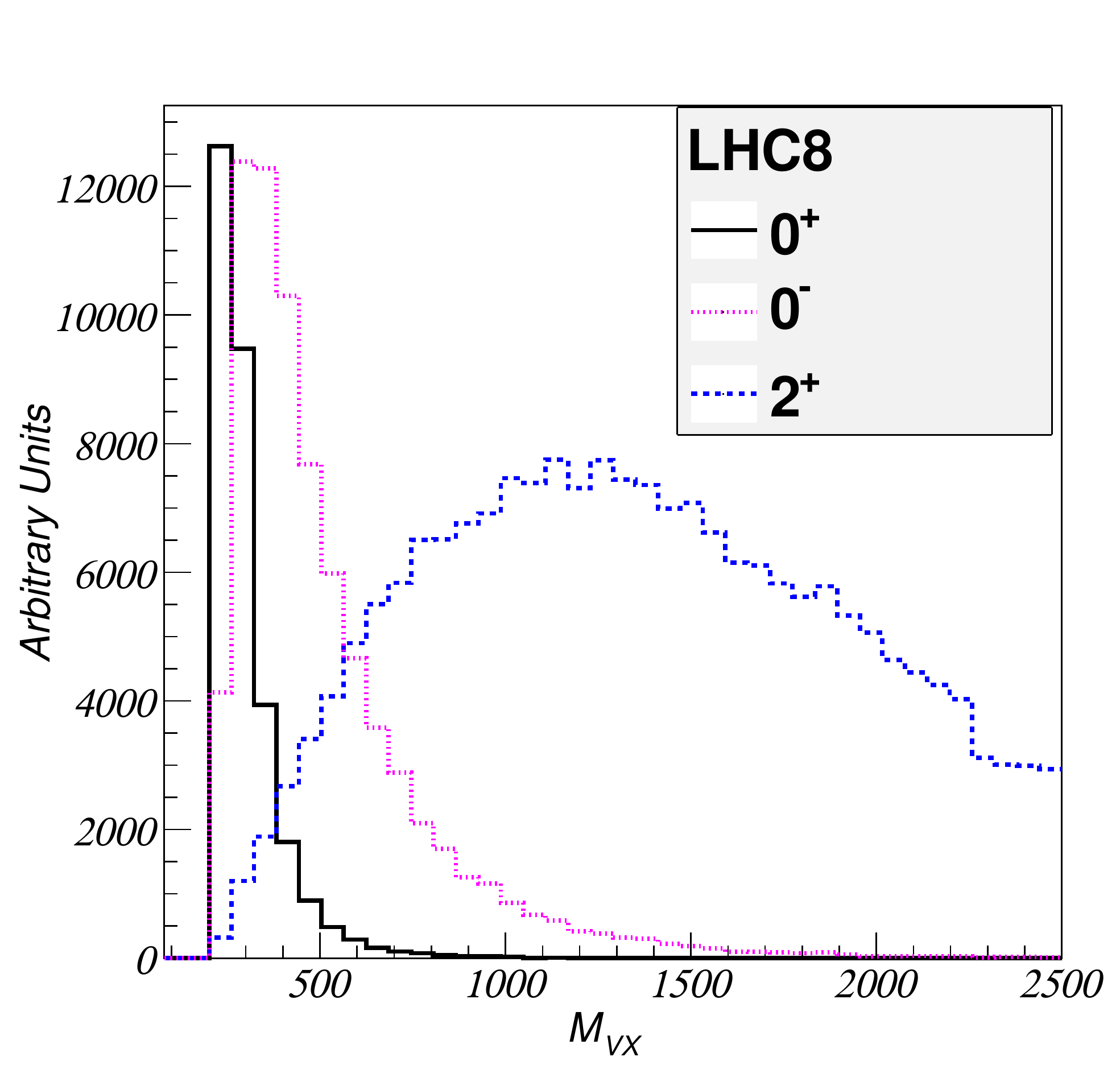}
\caption{\it Parton-level distribution of the $Z+X$ invariant mass for the $J^P = 0^+,0^-,2^+$ hypotheses in solid black, dotted pink and dashed blue lines respectively, at the Tevatron on the left and LHC at 8 TeV on the right. }
\label{fig:beta1or3}
\end{figure}

\begin{figure}[h!]
\centering
\includegraphics[scale=0.27]{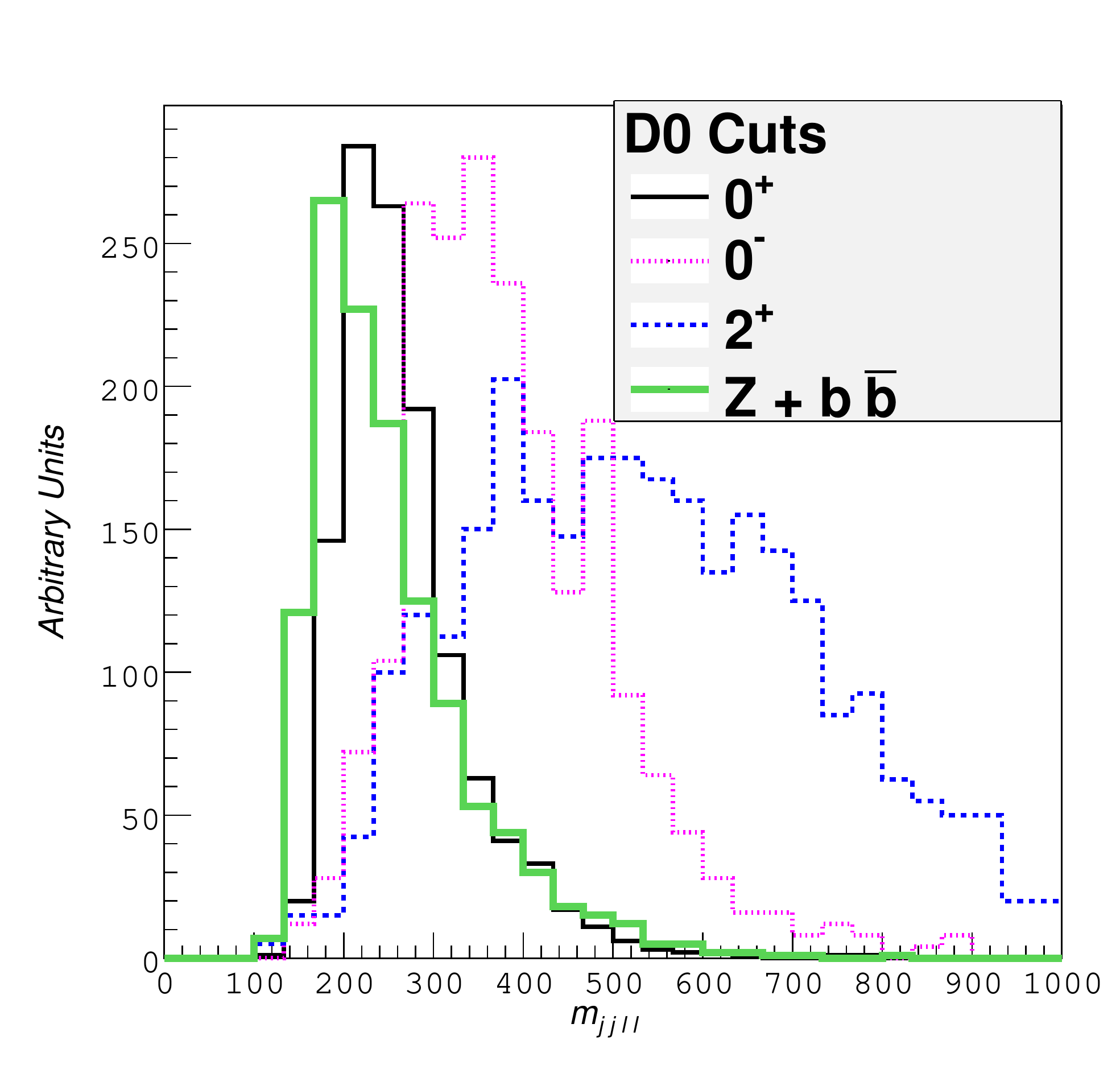}
\includegraphics[scale=0.27]{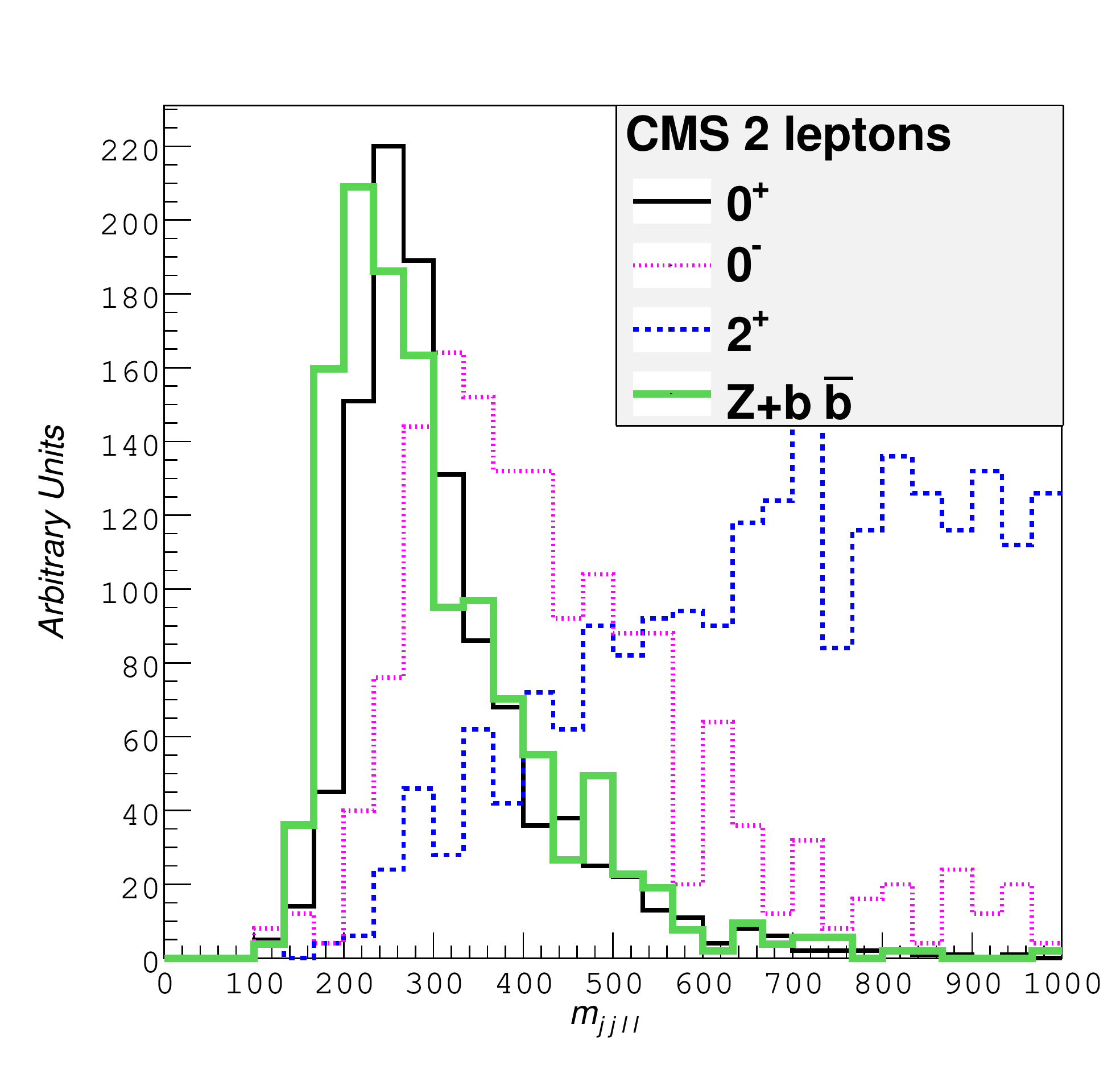}
\caption{\it Distribution after cuts of the $Z+X$ invariant mass in the 2-lepton channel for the $J^P = 0^+,0^-,2^+$ hypotheses in solid black, dotted pink and dashed blue lines respectively, for D0 on the left and CMS at 8 TeV on the right. The $Z+b\bar{b}$ background is shown in green.}
\label{fig:BG}
\end{figure}

On Fig.~\ref{fig:beta1or3} the different $Z+X$ invariant mass distribution at the parton level for $J^P$ spin-parity hypotheses $0^+, 0^-$ and $2^+$ is plotted in solid black, dotted pink and dashed blue lines respectively. The simulation was performed in MadGraph, with more information and details of the model implementation in Ellis et al.~\cite{fasttrack}. The figure clearly shows a difference between the scenarios at both the LHC and Tevatron, which can be understood by the production being an $s$-wave process for the $0^+$ case, whereas for $0^-$ this is $p$-wave and $2^+$ contains $d$-wave contributions. This difference survives the experimental cuts as illustrated for example in Fig.~\ref{fig:BG} for the 2-lepton channel at D0 and CMS, with the the background from $Z+b\bar{b}$ in green. 

The 0-,1- and 2-lepton channels at D0, CDF, CMS and ATLAS are simulated with experimental cuts (assuming no backgrounds) with the separation significance between two hypotheses $A$ and $B$ quantified by a log-likelihood ratio 
\begin{equation}
\Lambda = -2 ln \left( \frac{\mathcal{L}_A}{\mathcal{L}_B} \right)	\quad ,
\end{equation}
where the likelihood for a spin hypothesis $s$ is obtained by multiplying the probability distribution function
\begin{equation}
\mathcal{L}_s = \prod_i pdf_s(x_i)
\end{equation}
for each event $x_i$ in a `toy' simulation. Running a set of toys for the number of events after cuts expected for each experimental analysis generates a distribution in our test statistic $\Lambda$. Quantifying the separation significance between the two distributions in numbers of $\sigma$, we list in Table \ref{table:separationsignificance} the results for the various hypotheses in each search category, as well as their combination. 

\begin{table}[h!]
	\center
	\begin{tabular}{ | c | c | c | c | c |}
		\hline
		 Experiment & Category & Hypothesis A & Hypothesis B & Significance in $\sigma$ \\ \hline
		CDF & 0l & $0^+$ & $2^+ (0^-)$ & 3.7 (1.3)    \\ \hline
		 & 1l & $0^+$ & $2^+ (0^-)$ &  2.5 (1.0)    \\ \hline
		 & 2l & $0^+$ & $2^+ (0^-)$ & 1.4 (0.78)   \\ \hline
		 & ~Combined~ & $0^+$ & $2^+ (0^-)$ & 4.8 (1.6)   \\ \hline\hline
		D0 & 0l & $0^+$ & $2^+ (0^-)$ & 3.5 (1.2)   \\ \hline
		 & 2l & $0^+$ & $2^+ (0^-)$ & 1.8 (1.2)     \\ \hline
		& Combined & $0^+$ & $2^+ (0^-)$ & 4.0 (1.6)  \\ \hline\hline
		 ATLAS & 2l & $0^+$ & $2^+ (0^-)$ & 2.4 (1.1)    \\ \hline
		CMS & 2l & $0^+$ & $2^+ (0^-)$ & 2.3 (0.70)     \\ \hline
	\end{tabular}
	\caption{\it Estimated separation significance between different $J^P$ hypotheses at the Tevatron and LHC.} 
	\label{table:separationsignificance}
\end{table}

Note that this is for the ideal case of a perfectly clean extracted signal, since we are assuming no backgrounds. However the $s/b$ ratio at the Tevatron in the $H\to b\bar{b}$ channel is low enough to expect good sensitivity, and indeed a recent analysis of the invariant mass distribution by the D0 collaboration has excluded the spin $2^+$ hypothesis with graviton-like couplings with $3.1\sigma$ significance~\cite{D0}.

\section{Energy Dependence of Associated Production}
\label{sec:energy}

\begin{figure}
\begin{center}
\includegraphics[height=5.5cm]{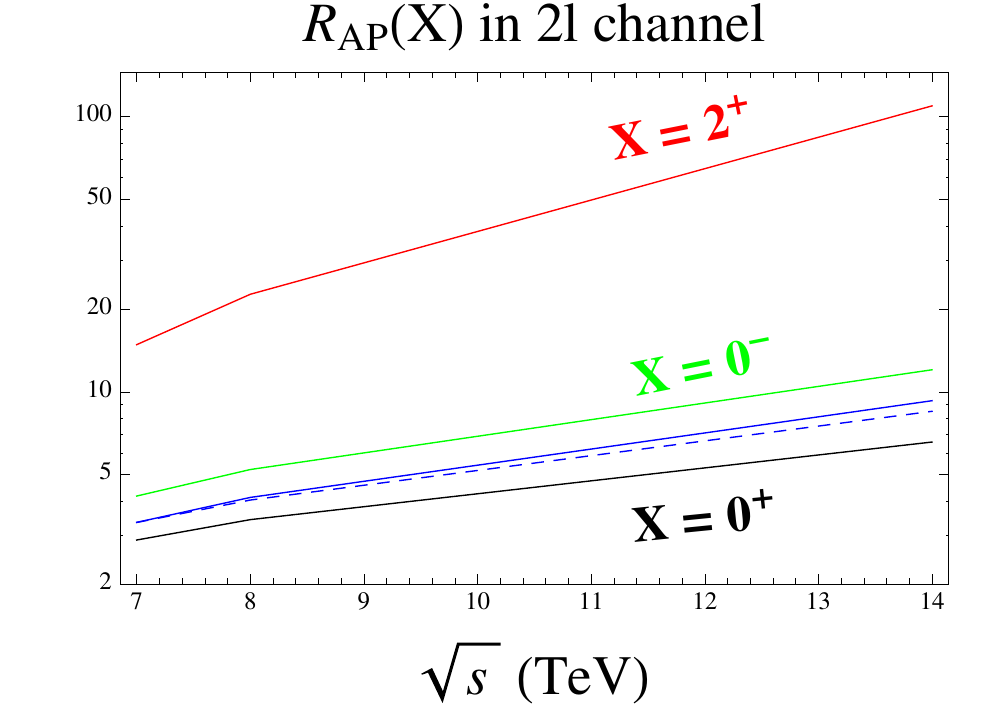}
\end{center}
\caption{\it
Ratio of the LHC signal strength relative to the Tevatron vs LHC energy for associated production $Z+X$ in the 2-lepton channel, after experimental cuts. The solid black, green and red line denote $J^P = 0^+,0^-$ and $2^+$ respectively, while the solid blue line is the dimension-6 contribution with $\epsilon_W=1$.}
\label{fig:Ranomalous}
\end{figure}

More information can be teased out of the fact that the Tevatron and LHC see a signal at different energies. The couplings of a pseudoscalar $A$ or graviton-like particle $G^{\mu\nu}$ have a different energy dependence to a SM Higgs due to their derivative couplings
\begin{equation}
\mathcal{L}_{0^-} \sim A F_{\mu\nu}\tilde{F}^{\mu\nu}	\quad , \quad
\mathcal{L}_{2^+} \sim G^{\mu\nu}T_{\mu\nu}	\quad . 
\end{equation}
The ratio of the signal strength in the 2-lepton channel at the LHC relative to the Tevatron is shown in Fig.~\ref{fig:Ranomalous} as a function of LHC energy for the $0^+, 0^-$ and $2^+$ hypotheses in black, green and red respectively. Note that this is including experimental cuts, with the 0-,1-lepton cases exhibiting an even stronger energy dependence. We refer the reader to Ellis et al.~\cite{secundafacie} for more details of the calculation. The signal expected at the LHC for a $2^+$ particle would be an order of magnitude larger than that of the Higgs. Since the observed signal strength is close to SM expectation within errors, this immediately provides evidence disfavouring such an interpretation.

For example at the 8 TeV LHC the ratio of the signal strength at the LHC relative to the Tevatron for the spin $2^+$ hypothesis divided by the same ratio for the SM $0^+$ hypothesis gives a double ratio of $\mathcal{R}_{2^+/0^+} = 7.4$. The observed double ratio extracted from the measured signal strength data yields $\mathcal{R}_{data} = 0.47 \pm 0.58$..

\begin{figure}
\center
\includegraphics[height=4.3cm]{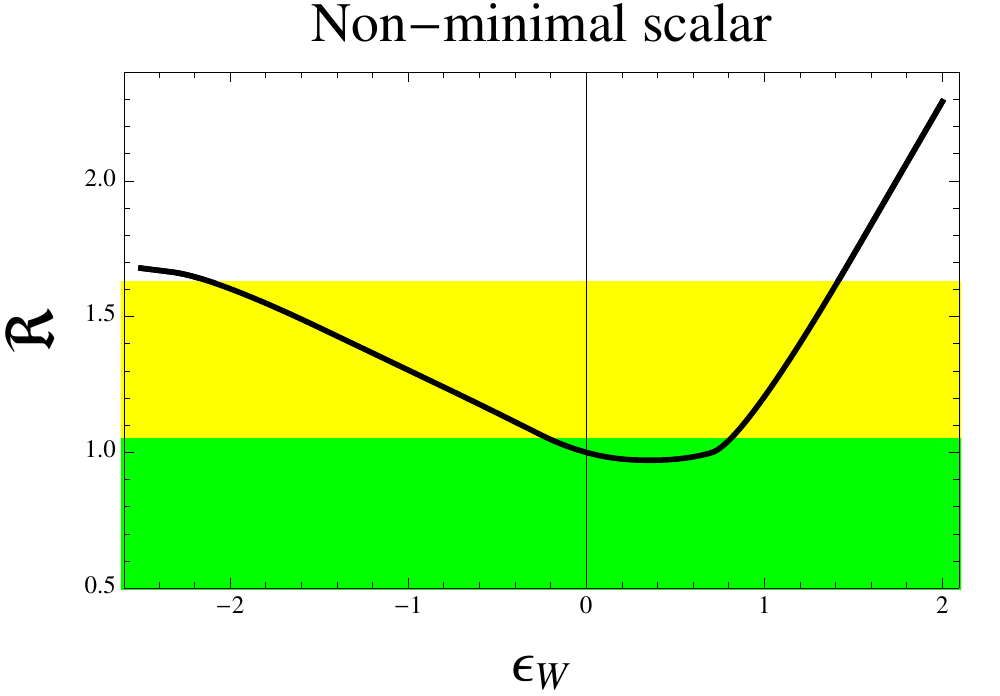}
\includegraphics[height=4.3cm]{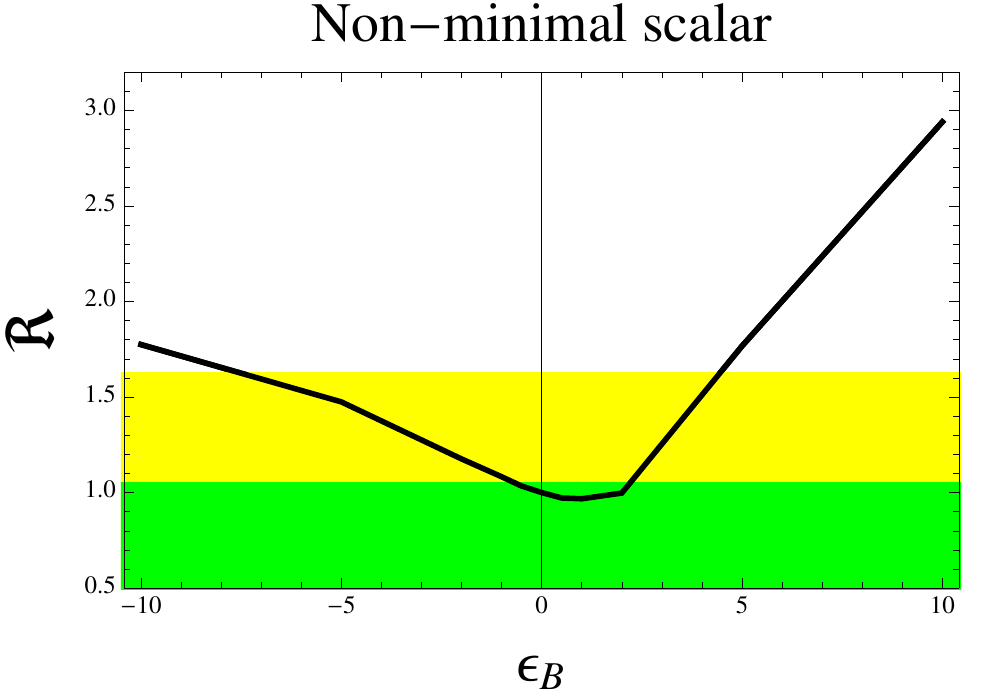}
\caption{
{\it
Double ratio $\mathcal{R}$ of the energy growth for dimension-6 operator contributions relative to that of a SM Higgs as a function of$\epsilon_W$ on the left and $\epsilon_B$ on the right. The experimental one and two sigma limits are shown as green and yellow bands.
} 
}
\label{eWR}
\end{figure}

This same method allows us to constrain the SM as an effective theory, in which higher-dimensional operators involving the Higgs contribute to measurable processes. In this case the derivative couplings of the Higgs doublet $\Phi$ in the operators
\begin{equation}
\mathcal{O}_W = (D_\mu \Phi)^\dagger \widehat{W}^{\mu\nu}(D_\nu\Phi)		\quad , \quad
\mathcal{O}_B = (D_\mu\Phi)^\dagger (D_\nu\Phi) \widehat{B}^{\mu\nu}		\quad ,
\end{equation}
are responsible for the energy dependence. Fig.~\ref{eWR} plots the double ratio $\mathcal{R}$ as a function of the operator coefficients $\epsilon_{W,B} \equiv f_{W,B} \frac{v^2}{\Lambda^2}$, with the experimental 1- and 2-$\sigma$ bands shown in green and yellow. The 95\% CL limits on $\epsilon_{W,B}$ can be read off as
\begin{equation}
\epsilon_W \in [-2.2,1.4]	\quad , \quad	\epsilon_B \in [-7.5, 4.4] 	\quad .
\end{equation}

\section{Conclusion }
\label{sec:conclusion}

The invariant mass distribution of the $V+H$ associated production mechanism provides a good discriminating variable to investigate further the properties of the newly-discovered particle. We have seen that this provides another handle on spin-parity measurements, with the D0 experiment at the Tevatron able to use this method to exclude spin $2^+$ at $99.9\%$ CL. We also showed how the energy dependence of this mode gives a complementary way of disfavouring a graviton-like particle or a pseudoscalar. Limits were placed on the coefficients of dimension-6 operators with derivative couplings that affect this energy dependence.

\section*{Acknowledgments}

The author thanks the organisers of the 25th Rencontres de Blois for the invitation to talk at an enjoyable and stimulating meeting, as well as John Ellis, Dae Sung Hwang and Veronica Sanz, with whom this work was carried out.

\end{document}